# Correlation Between Dopant Atom Evaporation Field and Measured Site Preference in Atom Probe Tomography

K. A. Hunnestad[1,2*], C. Hatzolgou[1], F. H. Danmo[1], Z. Yan[3], E. Bourret[3,4], F. Vurpillot[5], A. T. J. van Helvoort[6], S. Selbach[1], D. Meier[1,7,8*]

[1]Department of Materials Science and Engineering, Norwegian University of Science and Technology (NTNU), Trondheim, Norway

[2]Department of Physics, University of Oslo, Oslo, Norway

[3]Department of Physics, ETH Zurich, Zürich, Switzerland

[4]Materials Sciences Division, Lawrence Berkeley National Laboratory, Berkeley, CA, USA

[5]Univ Rouen Normandie, INSA Rouen Normandie, CNRS, Normandie Univ, GPM UMR 6634, F-76000, Rouen, France

[6]Department of Physics, NTNU, Trondheim, Norway

[7]Faculty of Physics and Center for Nanointegration Duisburg-Essen (CENIDE), University of Duisburg-Essen, Duisburg, Germany

[8]Research Center Future Energy Materials and Systems, Research Alliance Ruhr, 44780 Bochum, Germany

*Corresponding author: Kasper.hunnestad@fys.uio.no, dennis.meier@uni-due.de

**Complex oxides possess a wide range of electric, magnetic, and optical properties that can be precisely tuned by chemical doping. The atomic-scale analysis of the property-controlling dopants, however, becomes increasingly difficult towards low doping levels. Atom probe tomography (APT) offers chemical sensitivity and spatial resolution to image individual dopant atoms down to a few parts per million. To reliably extract such information, detailed knowledge about the atom-specific field evaporation processes is required. Here we demonstrate a first insight into the APT-measured atomic position of dopant atoms and the field evaporation conditions, using Zr-doped $ErMnO_3$ as a model system. Our analysis reveals a substantial preferential retention of both matrix and dopant atoms which strongly affects the dopant site determination and can lead to an incorrect interpretation. The retention effect is determined by intrinsic and extrinsic parameters, such as the dopant's evaporation field and concentration and the analysis temperature, respectively, as we explain based on field-evaporation simulations. Our results are important for the APT-based analysis of individual dopant atoms in solid systems and the understanding of field evaporation dynamics at the atomic level in general.**

## Introduction

Atom probe tomography (APT) is well suited for the investigation of doping and impurity atoms in complex oxides, combining high chemical sensitivity with sub-nanometer spatial resolution, such that even solute elements can readily be detected[1,2]. APT can give important atomic-level information on, for example, clustering, dopant association, and compositional fluctuations[3,4]. This capability is especially valuable for oxide semiconductors, where defects and subtle variations in dopant concentration can critically influence the materials' electronic responses and performance in devices[5–10]. Importantly for the investigation of more complex systems – where dopants can occupy different lattice sites and / or interstitial spaces, site-preferences of individual solute impurity atoms can be resolved [11], enabling direct observation of chemical anti-site defects, substitutions, interstitials, as well as defect complexes.

Reliable extraction of atomic-level information, however, strongly relies on detailed knowledge about the field evaporation of surface atoms, which is essential for the spatial reconstruction in APT[1,12,13]. Especially for materials where dopant and matrix atoms with completely different field evaporation behaviors coexist, special attention is required as elements with an above average evaporation field will evaporate delayed with respect to the expected evaporation sequence. This effect is referred to as preferential retention and is not covered in basic reconstruction protocols that assume only a single evaporation field for the entire system and, as a consequence, atomic positions are incorrectly calculated in the reconstructed volume[14–18]. An example is the observation of unexpected inter-atomic distances for layered systems with two or more elements[19,20]. In comparison to layered systems, where deviations from the expected periodic structure are comparatively easy to determine, signatures of atom-specific field evaporation processes in doped materials are much more subtle and harder to identify. Due to the low chemical concentration, dopant ions exert only minimal influence on the field evaporation dynamics and the reconstruction protocol of the atomic planes they are situated in[21]. Still, dopants with a different evaporation field are likely to follow a different evaporation sequence than their neighbors, as the local atomic neighborhood has a substantial effect on the evaporation sequence. For example, flat atomic terraces at the surface of the specimen can create chain effects where all atoms of one atomic plane evaporate simultaneously – or within a short time span[22,23], masking differences in the field evaporation strengths of dopants. Because of these evaporation-related effects, the information extraction at the level of individual dopant atoms remains a major challenge, including basic aspects such as site-preference and interatomic spacing.

Here, we study the relation between dopant-specific field evaporation and chemical mapping of individual impurity atoms. Using Zr-doped hexagonal erbium manganite ($ErMnO_3$) as model system, we measure the site-preference of the Zr-dopants under various experimental conditions and dopant concentrations. In contrast to the atomic position expected from the Shannon ionic radius and density functional theory calculations, our APT measurements suggest that the Zr dopants occupy the Mn-sites rather than the Er-sites. We explain this phenomenon based on preferential retention of Zr due to the high evaporation field compared to Er and Mn, corroborated by field evaporation simulations.

**Results and discussion**

Hexagonal manganites have been intensively studied with respect to their ferroelectricity[24], multiferroicity[25], and functional domain walls[26–28], and their atomic lattice structure[29–31] is well understood, which makes them an ideal model system for advanced doping-related studies. Most importantly for this work, it has already been demonstrated that the atomic planes are resolvable in APT, and that individual Ti dopant atoms in $Er(Mn,Ti)O_3$ can readily be resolved by APT along with their Mn-site preference[11,19]. Going beyond previous work, we now expand the work toward Zr-doped systems ($(Er,Zr)MnO_3$ with 0.04 at.% Zr concentration), where the dopant atoms are substantially larger. Thus, in contrast to the Ti atoms in $Er(Mn,Ti)O_3$, the Zr atoms in $(Er,Zr)MnO_3$ favor the Er-site[29,32] and different evaporation fields are expected. As a starting point, site-preferences of dopant atoms are often deduced by comparing their Shannon ionic radius to that of the lattice atoms of the host material[33]: $Ti^{4+}$ ($r_{Ti}$ = 0.51 Å, coordination number = 5) has an effective ionic radius comparable to $Mn^{3+}$ ($r_{Mn}$ = 0.58 Å) favoring the B-site as confirmed theoretically by density functional theory (DFT)[32] and experimentally by APT[11]. In contrast, $Zr^{4+}$ ($r_{Zr}$ = 0.78 Å, coordination number = 7) has a larger effective ionic radius and is expected to replace $Er^{3+}$ ($r_{Er}$ = 0.945 Å) on the A-site. Similarly to $Er(Mn,Ti)O_3$, DFT calculations support this effective-radius-based conclusion for $(Er,Zr)MnO_3$, [29] but remain to be verified experimentally (see Fig. 1(a),(b) for an illustration). Both $Ti^{4+}$ and $Zr^{4+}$ are donor dopants in $ErMnO_3$, and can be charge compensated by $Mn^{2+}$ at high temperature or low $pO_2$, or by interstitial oxygen, $O_i$, at low temperature and high $pO_2$[34]. The present work does not aim to distinguish these compensation mechanisms.

To determine the atomic position of the Zr-dopants, we perform APT measurements as described in Ref. [11] (see Methods for details). Figure 1(c) presents the 3D reconstruction considering all ionic species, whereas only the distribution of Zr is shown in Fig. 1(d). In general, the 3D data show no indication of secondary phases or clustering. The mass spectrum acquired from the dataset is shown in Fig. 1(e), where the $ZrO^{2+}$ peaks are clearly seen with no significant influence from other ionic species. Based on the APT data, the concentration can be estimated to about 0.033 ± 0.008 at.%, which is close to the nominal value of 0.04 at.% estimated during material synthesis.

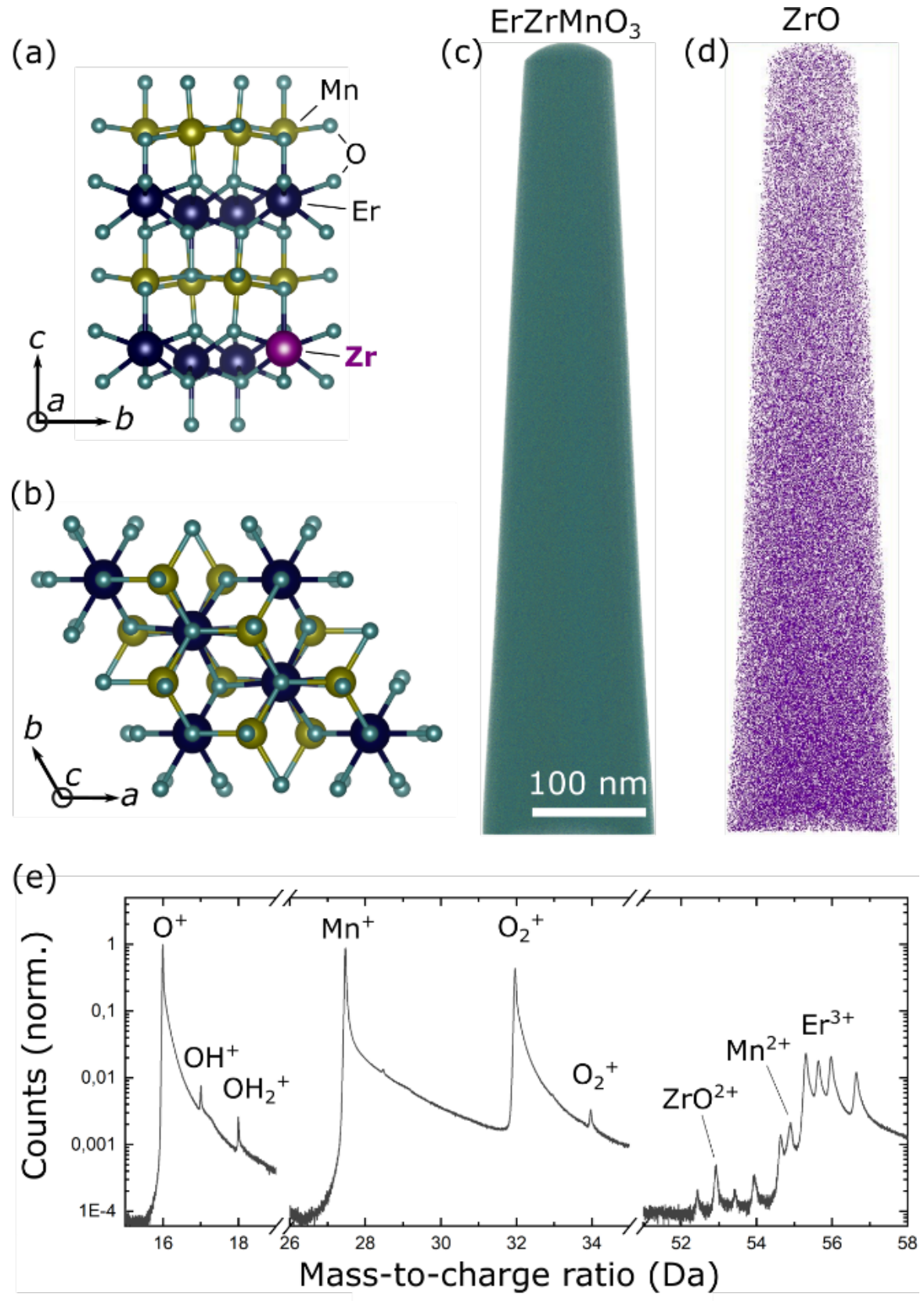

**Figure 1. Zr dopants in (Er,Zr)MnO$_3$.** (a,b) Unit cell viewing along the *a* and *c*-axis, respectively. In (a), a Zr dopant is depicted on the Er site as expected based on its Shannon ionic radius and DFT calculations. (c) 3D reconstruction from the APT analysis of (Er,Zr)MnO$_3$ with 0.04 at.% Zr concentration (all elements) (d) Same as in (c) showing only the Zr dopants (from the ZrO ionic species) in e. (e) Mass spectrum of (Er,Zr)MnO$_3$.

At this point, it is important to note that the samples for APT analysis are prepared so that the hexagonal *c*-axis points along the needle axis, such that the (001)- crystallographic pole forms in the center of the needle. The reason for orienting the samples this way is that in the crystallographic poles the spatial resolution is highest, facilitating the observation of atomic planes[11,19,22]. In Figure 2(a), a 3D image from the pole region is shown, including the Mn atoms (grey) to represent the matrix, as well as Zr dopants (purple). A projection parallel to the *c*-axis is shown in Fig. 2(b), where the atomic planes formed by the Mn are clearly visible. Regarding the position of the Zr dopant atoms within this limited volume, no clear trend is visible in this viewing direction, requiring a more detailed statistically representative analysis.

To obtain accurate and statistically significant information on the average Zr-dopant position, we calculate spatial distribution maps (SDMs) of the entire volume using Mn as a reference (Fig. 2(c)). The SDMs are calculated by summing up the differences in a single coordinate between two ionic species; here, the distance along the crystallographic *c*-axis (z-SDM)[35,36]. The position of the largest peak in the SDM then indicates the average distance of an ionic species away from the Mn planes. For the Mn-Mn SDM, the largest peak is at zero, as the same ionic species share atomic planes. In contrast, the Mn-Er SDM shows a peak at about 0.25 $d_{Mn}$ away from the center. This APT-measured value is significantly smaller than the distance expected from the established unit cell structure depicted in Fig. 1(a) (that is, 0.5 $d_{Mn}$), reflecting pronounced preferential retention[12]. The preferential retention of the Er is due to its higher evaporation field compared to Mn and consistent with previous APT experiments on $ErMnO_3$[19].

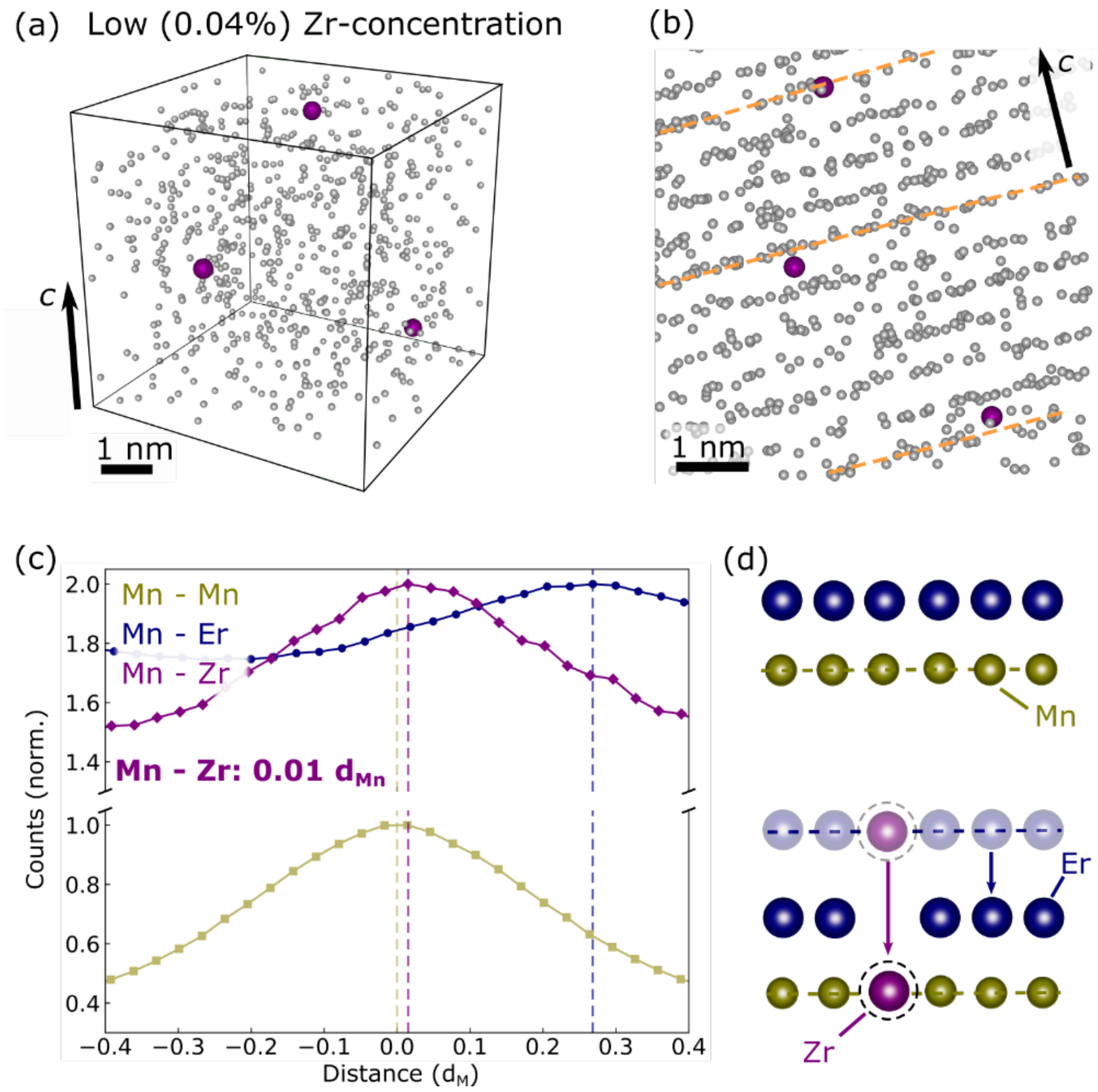

**Figure 2. Atomic-level analysis of Zr dopants.** (a) and (b) show sections from the (001) crystallographic pole from the APT dataset where atomic resolution is achievable. Only the Mn (grey) and Zr (purple) atoms are shown, and orange striped lines indicate the Mn planes adjacent to the Zr dopant. (c) SDM with the atomic positions of Er and Zr relative to Mn, where the x-axis is normalized to Mn plane-to-plane distance ($d_{Mn}$). The Zr peak (indicated by purple dashed line) is located on the Mn peak (yellow dashed line) instead of the Er peak (blue dashed line). This corresponds to the atomic positions sketched in (d), where Zr is positioned on the Mn site instead of the Er site.

Most noteworthy, we observe that the primary peak of the Mn-Zr SDM is located very close to zero. This result is surprising as it suggests a Mn site-preference for Zr, which is the opposite to the Er site-preference expected for Zr based on the mentioned Shannon ionic radius of Zr and DFT calculations[29]. This clear contradiction can be resolved accounting for the Zr evaporation field, which is not explicitly considered in the standard reconstruction approach used in APT. The default algorithm assumes a single evaporation field for the entire (Er,Zr)$MnO_3$ system. Consequently, elements that are low in concentration have only a minimal influence on the reconstruction and, hence, are more likely to be positioned on incorrect atomic planes during the reconstruction. It is also worth mentioning that large lateral displacements in the initial field evaporation can also play a role here, but these are assumed to be second-order effects in this work[20]. The experimental data thus leads us to the conclusion that Zr has a larger evaporation field than Er (which is larger than Mn), so that Zr is retained on the surface after evaporation of the Er planes. Note that the standard field desorption model suggests a higher evaporation field of pure Mn than pure Zr (30 V/nm compared to 28 V/nm)[37,38], and is insufficient here. The results of the Mn-Zr peak at zero in the SDM are presented in Fig. 2(c), and a resulting reconstruction is sketched in Fig. 2(d). This result strikingly shows the importance of the dopant's evaporation field and the need for adequate testing when APT is applied to evaluate site-preferences in lightly doped oxides and impurity atoms in crystalline solids in general.

To corroborate our conclusion and highlight the key role the evaporation field of dopants plays for the APT analysis, we next vary both the defect concentration (intrinsic) and temperature (extrinsic), i.e., primary material- and experiment-related factors that control the evaporation dynamics and evaluate respective changes in the SDMs. If correct, a higher dopant concentration is anticipated to increase the "apparent average" evaporation field of the Er plane (including the Zr dopants) and the relatively high-field effect of Zr should become less significant. Furthermore, differences in evaporation field between various ionic species are known to change with temperature[39]; specifically the difference between the Er and Mn ionic species is reduced at lower temperatures[19]. Thus, temperature variations are expected to directly affect the evaporation field differences between the dopant and the matrix. In the following, respective doping- and temperature-dependent test experiments are discussed.

Figure 3(a) presents SDMs gained on (Er,Zr)$MnO_3$ with a 50 times higher Zr concentration (2 at. % Zr, see Methods for details) than in Fig. 2(c). On the one hand, the SDMs calculated from the (001)-pole reveal that the Mn-Er primary peak is closer to zero than in Fig. 2(c), indicating a much more pronounced preferential retention of Er for the higher doping level. This observation is in line with a higher Zr evaporation field compared to Er, leading to an increase of the average Er/Zr-plane evaporation field as the density of Zr in the Er-planes increases, increasing the preferential retention of Er planes overall. On the other hand, we observe that the position of the Mn-Zr SDM peak is no longer centered around zero, but has shifted towards the Mn-Er peak, while still reflecting pronounced retention effects. Such retention effects can in general be suppressed by reducing the lower laser pulse energy, which leads to a lower peak temperature, $T$, during pulsing and, hence, smaller evaporation field differences[19]. Corresponding SDMs gained at lower laser pulse energy are shown in Fig. 3(b). Consistent with our proposed evaporation scenario and the theoretically predicted Er site-preference of Zr, retention effects are found to be reduced at lower pulsing energy and the peak for the Mn-Zr distance is clearly shifted away from zero. We note that although the applied increase in doping and reduction in peak temperature does not fully suppress retention-related artefacts, the observed trends are clear, demonstrating a close correlation between the evaporation field of the dopant atoms, their concentration, analysis temperature, and the resulting positional information.

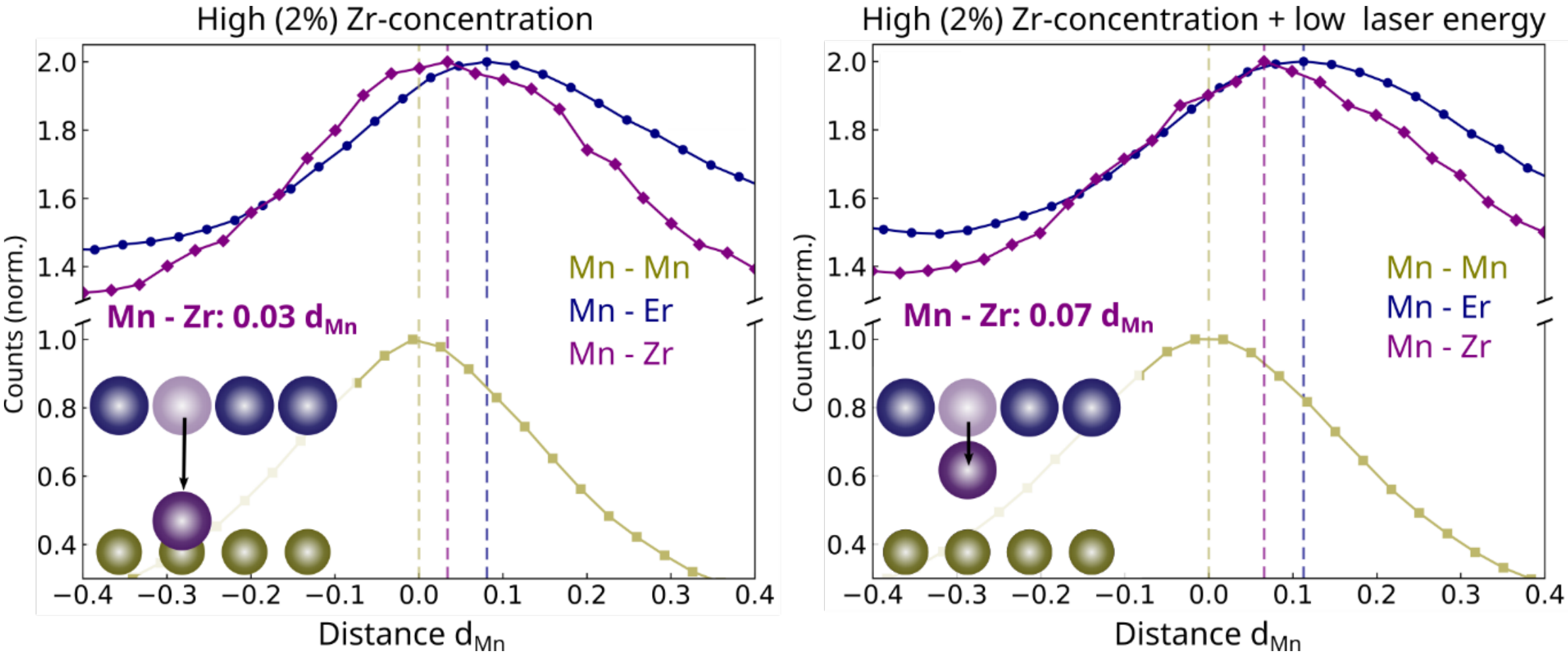

**Fig. 3. Obtaining accurate site-preference of Zr.** (a) SDM showing the positions of Er and Zr relative to Mn for (Er,Zr)$MnO_3$ with a high nominal Zr concentration of 2 at.%. There is a strong tendency for preferential retention, causing the atomic planes to overlap, although the Zr peak (purple dashed line) can be seen to be shifted away from the Mn peak (yellow dashed line) and lie closer to the Er peak (blue dashed line). (b) SDM of the same sample analyzed at low laser pulse energy and low temperature, *T*, where the Zr and Er peaks are located further away from the Mn peak.

On a qualitative level, this dependence can be modelled by field-evaporation simulations, allowing the evaporation sequence to be tracked directly, starting from a well-defined initial structure without the complexities of a 3D reconstruction. For this purpose, we model a simplified crystal with alternating Er and Mn planes, introducing 5 at.% Zr dopants on the Er site (1 at.% nominal concentration). Field-evaporation simulations are carried out as described in Refs. [23,40–43], and the results for four representative cases are shown in Fig. 4. Each histogram in Fig. 4 displays the dopant evaporation sequence relative to the Er plane (defined as zero). Because the evaporation order in APT is directly linked to the depth coordinate, we convert this to a distance for clarity. The histogram position indicates whether the dopants evaporate earlier (advanced evaporation) or later (delayed evaporation) than the Er plane, depending on their evaporation field relative to the evaporation field of Er ($F_{Er}$). For a reduced Zr evaporation field ($F_{Zr} < F_{Er}$, low-field dopant, Fig. 4a), the distribution shifts to the right, indicating earlier evaporation. When $F_{Zr} = F_{Er}$ (Fig. 4b), the distribution centers on the Er plane. For $F_{Zr} > F_{Er}$ (Fig. 4c,d), the distribution shifts to the left, showing delayed evaporation, characteristic of preferential retention after the Er plane has evaporated.

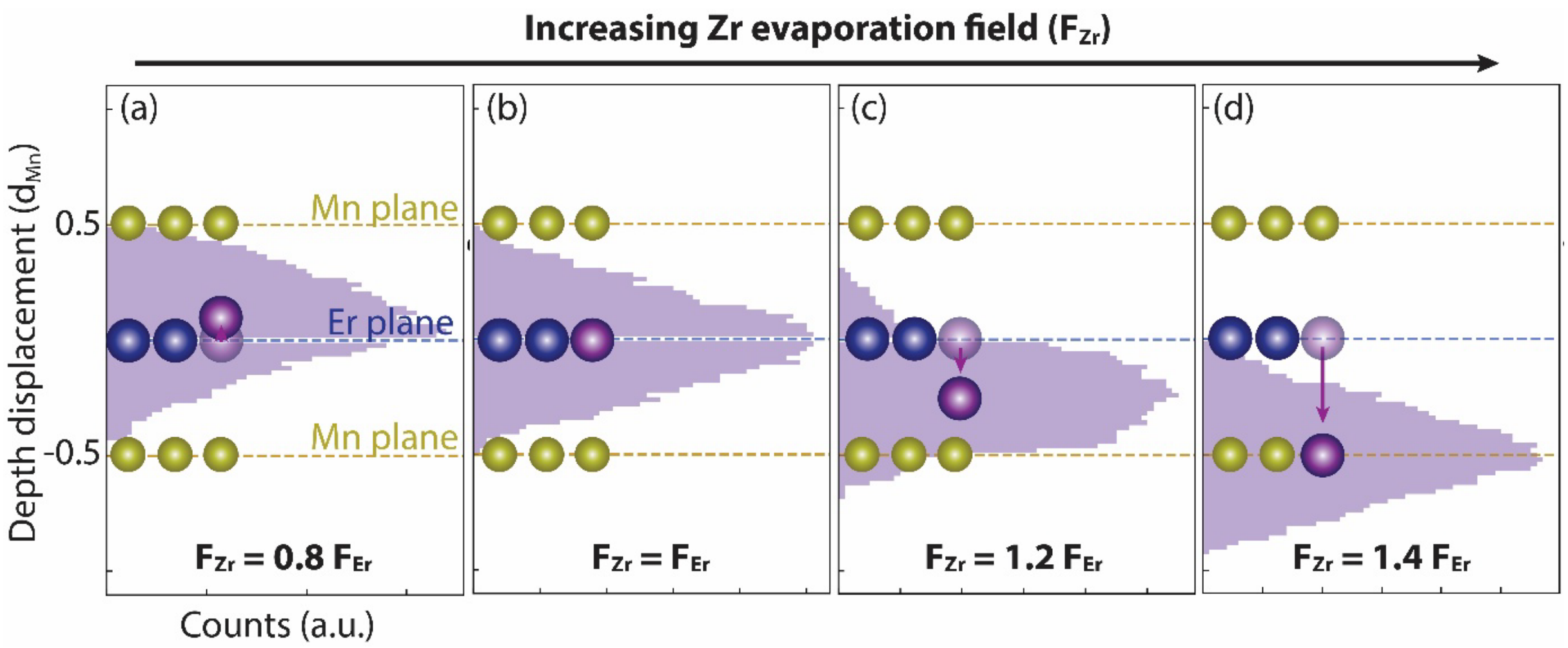

**Fig. 4**: **Field evaporation sequence with varying dopant evaporation field**. (a-d) Histograms of the evaporation sequence of the Zr dopants, with an evaporation field ($F_{Zr}$) of 0.8, 1, 1.2 and 1.4 times that of Er. (a) Advanced evaporation of the Zr dopant. (b) Normal evaporation sequence of the dopant. (c) and (d) preferential retention of Zr, leading to delayed evaporation. The y-axis is the depth displacement from its original atomic plane. Schematic representations of the apparent dopant position is shown with each panel.

To provide a more comprehensive view of how the Zr evaporation field affects both the evaporation sequence and reconstructed position, we performed a full series of simulations covering $F_{Zr}$ values from 0.5 $F_{Er}$ to 1.8 $F_{Er}$ (Fig. 5). The overall trend fits the representative cases in Fig. 4, that is, low-field dopants evaporate earlier than their host atoms, producing a positive depth displacement, whereas high-field dopants evaporate later, yielding a negative displacement. Notably, the behavior is asymmetrical. For low-field dopants, the depth displacement rapidly saturates at a small positive value. In contrast, for high-field dopants, no such limit is observed — the preferential retention effect continues to increase with the evaporation field.

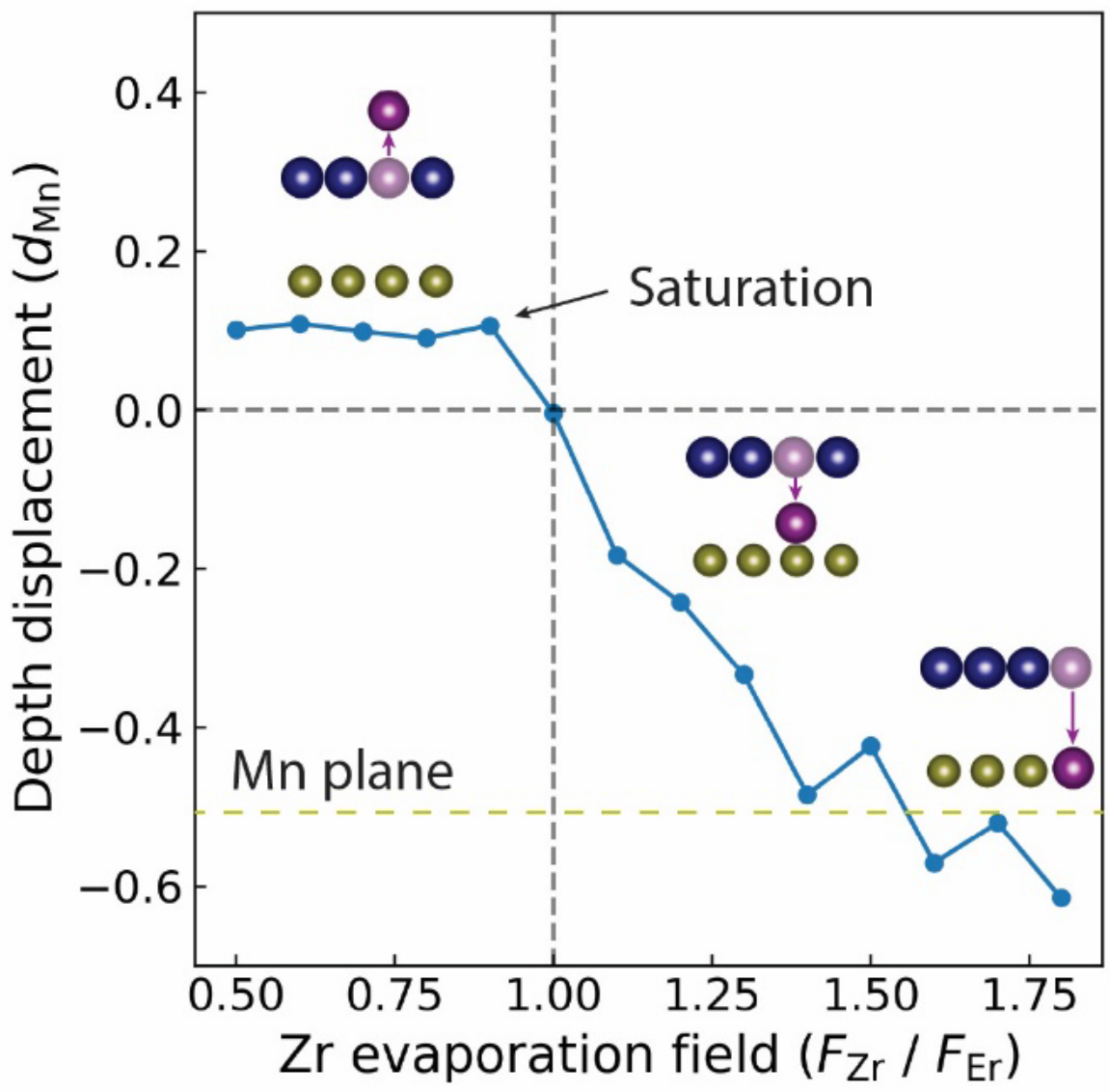

**Fig. 5**: **Preferential retention effect as a function of dopant evaporation field**. The plot shows the average value extracted from Fig. 4 for different Zr evaporation fields ($F_{Zr}$). Negative values on the y-axis indicate preferential retention, whereas positive values indicate an advanced evaporation. The schematics illustrate the consequences of the evaporation sequence and resulting depth displacement in the final reconstruction. Blue atoms are Er, yellow are Mn.

Figure 5 shows that in extreme cases the dopants can be retained beyond the subsequent atomic plane. This effect is shown visually in Fig. 6 for a dopant with an evaporation field of $F_{Zr} = 1.8\ F_{Er}$, where each panel represents the evaporation of a single atomic plane. The dopant is not just retained after the evaporation of its neighboring Er atoms, it is also retained after the evaporation of the subsequent Mn plane. Interestingly, the retained dopant causes delays to the evaporation of the Mn atoms situated below the Zr dopant, demonstrating a much more complex evaporation behavior than documented previously. Thus, although only one dopant atom is present, substantial distortions arise in the spatial reconstruction of its environment. In other words, even a solute species can alter the matrix evaporation sequence and potentially influencing the overall APT analysis.

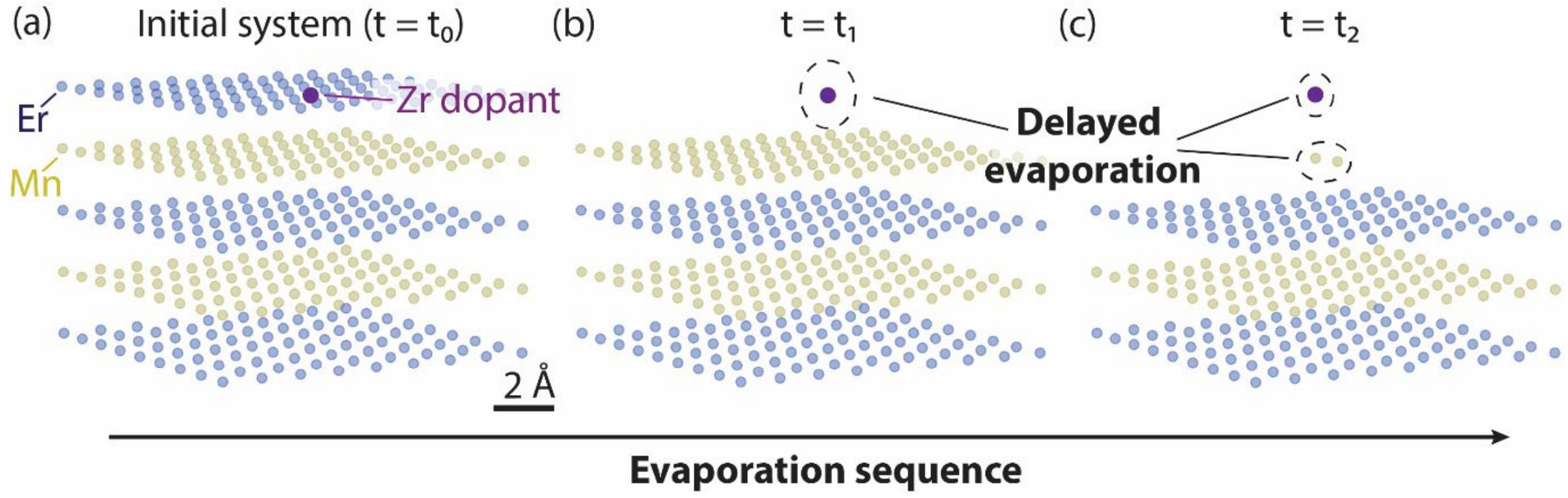

**Fig. 6**: **Visualization of the evaporation sequence for dopants with a large evaporation field**. (a) The initial system used in the simulation, with a single Zr dopant placed within the top Er atomic plane with an evaporation field $F_{Zr} = 1.8\ F_{Er}$. (b) and (c) shows the system after some time ($t_2 > t_1 > t_0$). In (b), the top atomic plane is evaporated, leaving behind the retained Zr dopant atom, whereas in (c), the dopant atom is still retained after the subsequent Mn atomic plane has been evaporated.

## Conclusion

In this work, we investigated how differences in evaporation field between dopant and matrix atoms influence the apparent atomic positions of solute dopants in APT, using Zr-doped $ErMnO_3$ as a model system. Although Zr is known to substitute Er atoms on the A-site, conventional APT reconstruction falsely places the dopants on the Mn planes. By combining systematic experiments with field-evaporation simulations, we demonstrate that this wrongly assigned site preference originates from preferential retention of high-field dopants during field evaporation, resulting in systematic reconstruction artefacts. The magnitude of this effect is governed by the relative evaporation field of the dopant, its concentration, and the experimental analysis conditions, with higher dopant concentrations and lower laser pulse energies reducing the apparent displacement. Furthermore, our simulations reveal that sufficiently large evaporation-field differences not only delay the evaporation of the dopant itself but also perturb the evaporation sequence of neighboring matrix atoms, leading to local distortions of the reconstruction. The findings establish a close relationship between dopant evaporation fields and the measured site preference in APT, requiring adequate control experiments to reliably extract atomic positions of solute atoms. The work highlights the fundamental importance of considering evaporation-field differences between dopants and the matrix for the reconstruction, which is of importance for the interpretation of APT data gained on doped complex oxides and crystalline systems in general.

## Acknowledgements

The Research Council of Norway (RCN) is acknowledged for the support to the Norwegian Micro- and Nano-Fabrication Facility, NorFab, project number 295864, the Norwegian Laboratory for Mineral and Materials Characterization, MiMaC, project number 269842/F50, and the Norwegian Center for Transmission Electron Microscopy, NORTEM (197405/F50). K.A.H. and D.M. thank the Department of Materials Science and Engineering at NTNU for direct financial support. D.M. acknowledges funding from the European Research Council (ERC) under the European Union's Horizon 2020 research and innovation program (Grant Agreement No. 863691). D.M. thanks NTNU for support through the Onsager Fellowship Program and NTNU Stjerneprogrammet.

## Methods

**Materials:** Single crystals of $Er_{0.998}Zr_{0.002}MnO_3$ are grown with the pressurized floating-zone method, oriented by Laue diffraction, and then cut to achieve flat polar surfaces[44]. Polycrystalline $Er_{0.9}Zr_{0.1}MnO_3$ samples were prepared through solid-state synthesis. Stoichiometric amounts of binary oxides ($Er_2O_3$, $ZrO_2$, $Mn_2O_3$) were thoroughly mixed and pressed into pellets, which were fired in air at 1300 °C for 1 h. The fired pellets were crushed to powders, re-pressed into pellets, and then sintered in air at 1400 °C for 10 h. Specimens for APT were prepared from lapped & polished bulk specimens using a Thermo Fisher Scientific G4 DualBeam UX Focused Ion Beam (FIB) analogous to the procedure described in Ref.[45].

**APT:** To analyze the APT specimens and collect data for analysis, a Cameca LEAP 5000XS was used, operating in laser pulsing mode. The initial APT analysis in Figure 1 and 2 was conducted in laser pulsing mode with a laser pulse energy of 5 pJ, a pulse frequency of 250 kHz and a base temperature at 25 K. Detection rate was set to 1 %, meaning that on average 1 atom is detected every 100 laser pulses. For the APT analysis in Figure 4(a), the laser energy was instead set to 10 pJ, the frequency was set to 500 kHz, and the detection rate was set to 4 %, while in Figure 4(b) the laser energy was lowered to 0.5 pJ. To reconstruct the raw data into 3D datasets, the software Cameca IVAS 3.6.12 was used. The radial evolution was estimated from the voltage profile. SDM analysis was done using the Norwegian Atom Probe App (NAPA) software, developed in MATLAB®. All SDM are calculated normal to the atomic planes to optimize signal-to-noise ratio. For ranging the $ZrO^{2+}$ ionic species, only the two lightest isotopes were used, as the heavier isotopes partly overlaps with $Er^{3+}$.

**Field evaporation simulation**: The simulation model used in this study has been developed at the Groupe de Physique des Materiaux (GPM) by Vurpillot et al. More details of this model and its applications can be found in Refs. [23,41–43]. To reproduce the studied system as faithfully as possible, we have simulated the field evaporation of a successive stack of atomic planes of Er and Mn (without the oxygen). The structure is simplified as a simple cubic structure. To account for the thermally active process of evaporation, a Monte-Carlo type algorithm chooses the atom to evaporate according to its evaporation rate. Given our previous observations on $ErMnO_3$[11,19], namely the preferential retention of Er atomic planes, the evaporation field of Er atoms ($F_{Er}$) are therefore higher than that of Mn ($F_{Mn}$) and is thus set equal to $1.2 \times F_{Mn}$. Subsequently, we considered that 5% of the Er sites were occupied by Zr atoms, thus corresponding to a Zr doping of 2.5 at.%, which corresponds to 1 at.% when including oxygen. Since the evaporation field of Zr is not precisely known, we have varied its relative evaporation field from 0.5 to 1.8 times that of Er ($F_{Er}$). The histograms are created directly from the field evaporation sequence from the field evaporation simulations using only the central 001-pole. For converting the x-axis from sequence to depth, a function that relates depth to atomic sequence was established from the reconstructed simulation, assuming a linear trend.